# An Energy-Efficient Configurable Lattice Cryptography Processor for the Quantum-Secure Internet of Things

Utsav Banerjee[1], Abhishek Pathak[2], Anantha P. Chandrakasan[1]
[1]Massachusetts Institute of Technology, Cambridge, MA
[2]Indian Institute of Technology Delhi, New Delhi, India

Modern public key protocols, such as RSA and elliptic curve cryptography (ECC), will be rendered insecure by Shor's algorithm [1] when large-scale quantum computers are built. Therefore, cryptographers are working on quantum-resistant algorithms, and lattice-based cryptography has emerged as a prime candidate [1]. However, high computational complexity of these algorithms makes it challenging to implement lattice-based protocols on resource-constrained IoT devices which need to secure data against both present and future adversaries. To address this challenge, we present a lattice cryptography processor with configurable parameters which enables up to two orders of magnitude energy savings and 124k-gate reduction in system area through architectural optimizations. This is also the first ASIC implementation which demonstrates multiple lattice-based protocols proposed in Round 1 of the NIST post-quantum standardization process.

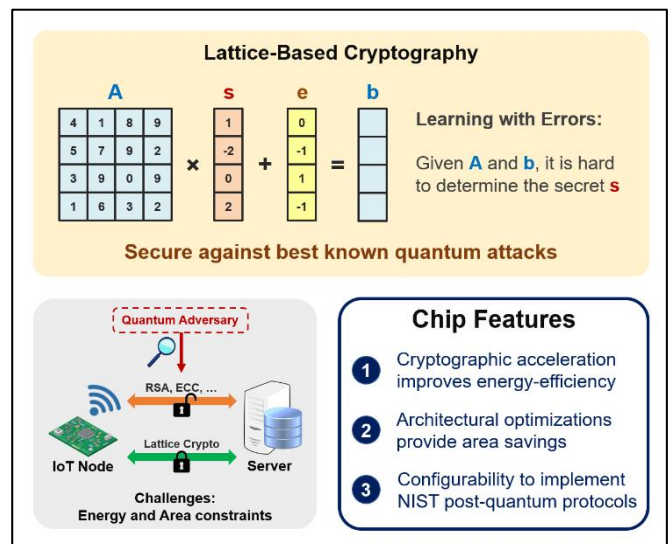

Fig. 1: Quantum-resistant security for IoT networks – lattice-based cryptography, challenges and proposed hardware solutions.

Fig. 1 provides an overview of the "Learning with Errors" (LWE) problem which forms the basis of several lattice-based schemes. The LWE hardness assumption states that it is computationally difficult to determine the secret vector $s$, given the matrix $A$ and the vector $b = As + e$, where all arithmetic is modulo a small integer $q$, and $s$ and $e$ are short vectors sampled from a discrete distribution. This hardness is preserved even in the presence of quantum adversaries. The two most commonly used variants of LWE are Ring-LWE and Module-LWE, which operate on polynomials instead of vectors for efficiency, both of which can be accelerated using our processor.

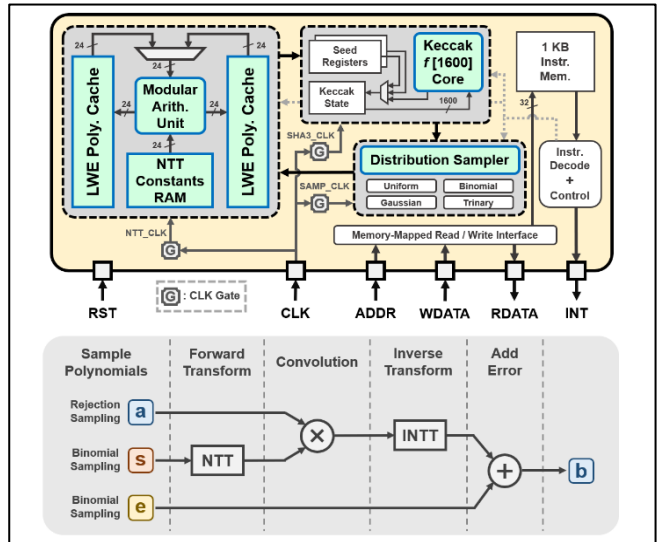

Fig. 2: System diagram along with overview of a typical Ring-LWE computation.

Fig. 2 shows the system block diagram, along with details of a typical Ring-LWE computation. A 24KB LWE Cache interfaces with a modular arithmetic unit to perform polynomial operations including the number theoretic transform (NTT). An energy-efficient Keccak-f[1600] core, used for hashing and pseudo-random number generation (PRNG), drives the discrete distribution sampler. The LWE cache, the Keccak core and the sampler have dedicated clock gates which can be independently configured for fine-grained power savings. The processor is equipped with a 1KB instruction memory which can be programmed with custom instructions to implement various lattice-based algorithms. Two most important computations required in all protocols are sampling and convolution. The polynomials are generated, or "sampled", either uniformly through rejection sampling or from a discrete distribution, typically binomial, with a carefully chosen standard deviation. Computing convolution of two polynomials involves transforming to the NTT domain followed by coefficient-wise multiplication and an inverse transform.

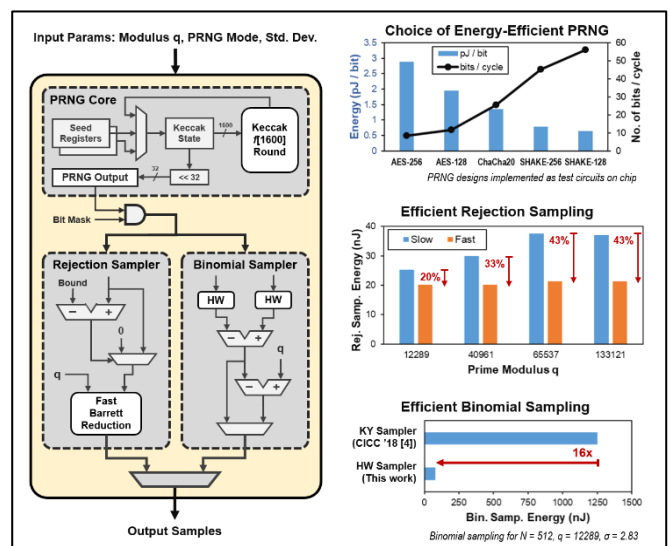

Fig. 3: Architecture of discrete distribution sampler with efficient PRNG and sampling.

The hardness of the LWE problem is directly related to the statistical properties of the sampled polynomials. This makes an accurate and efficient sampler a critical component of any lattice crypto implementation. Sampling accounts for about 70% of the computational overhead in software implementations of lattice-based protocols [2]. Fig. 3 describes an energy-efficient discrete distribution sampler which reduces this overhead and provides up to two orders of magnitude energy savings over assembly-optimized software. Samplers post-process pseudo-random bit strings to generate numbers from a specified distribution, thus making an efficient PRNG a key requirement for energy savings. Hardware implementations of three standard PRNGs with full data-path architectures were profiled on our test chip, and SHA-3 (SHAKE) was observed to be 2x and 3x more energy-efficient than ChaCha20 and AES respectively. Therefore, our PRNG consists of a 24-cycle 34k-gate Keccak-f[1600] core which can be configured in different SHA-3 modes and consumes 0.89 nJ per round. Our Keccak core processes its 1600-bit state in parallel, thus avoiding expensive register shifts and multiplexing required in serial architectures. The associated area overhead is very small, since the PRNG accounts for only 9% of the total processor area. Rejection sampling for primes with high rejection probability can be a bottleneck in LWE-based protocols. For faster rejection sampling, the rejection bound is set as a multiple of the prime modulus $q$ [3] followed by Barrett reduction, providing up to 43% energy savings compared to conventional rejection. Our binomial sampler takes two $k$-bit chunks ($k \leq 32$, configurable) from the PRNG and computes the difference of their Hamming weights (HW) to generate a sample with standard deviation $\sigma = \sqrt{(k/2)}$. This method is 16x more energy-efficient than the conventional Knuth-Yao (KY) sampler [1, 4], and is also constant-time, thus eliminating potential timing side-channels.

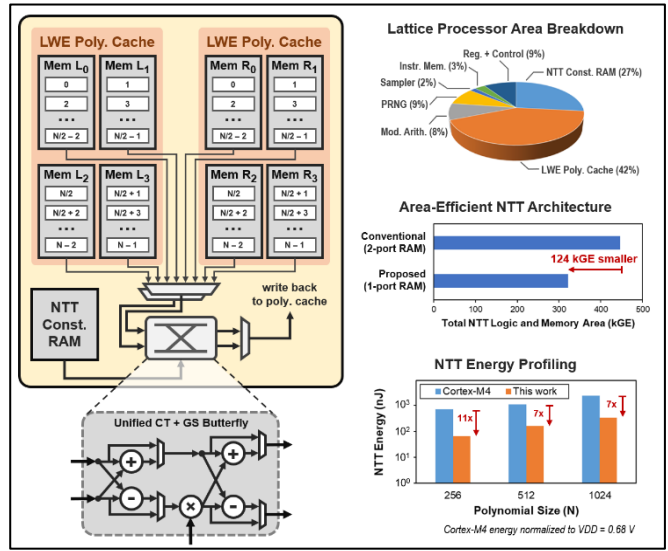

Fig. 4: Proposed single-port RAM-based area-efficient NTT architecture with processor area breakdown and NTT energy profiling.

Polynomial operations, such as NTT and convolution, account for about 30% of the computations. However, the associated memory and logic together occupy more than 75% of the total hardware area. Hardware architectures for NTT, first proposed in [1], consist of SRAM banks for storing polynomials along with a modular arithmetic unit to perform the butterfly computations. These memories are typically implemented using two-port [1] or four-port [4] RAMs, which can pose large area overheads in resource-constrained devices. To reduce this area, we implement the constant geometry NTT [5] and split each polynomial among 4 single-port RAMs, as shown in Fig. 4. Regular memory access patterns of the constant geometry NTT allow butterfly inputs and outputs to ping-pong between these single-port RAMs without any read or write hazards. This NTT architecture provides ~124k-gate area savings compared to the traditional approach, while still having enough memory to accommodate multiple polynomials required in lattice-based algorithms. The constant factors $\omega$ and $\psi$ used in NTT-based negative-wrapped convolution are related as $\omega = \psi^2$ and $\omega^{-i} = \omega^{N-i}$, which is used to compress pre-computed tables stored in the NTT Constants RAM by 38%. The butterfly, with a 24-bit data-path and configurable modulus $q$, is implemented as a unified Cooley-Tukey (CT) + Gentleman-Sande (GS) structure, which eliminates the need for expensive bit reversals. The multiplier and adder/subtractor in the butterfly are re-used for coefficient-wise modular operations on polynomials.

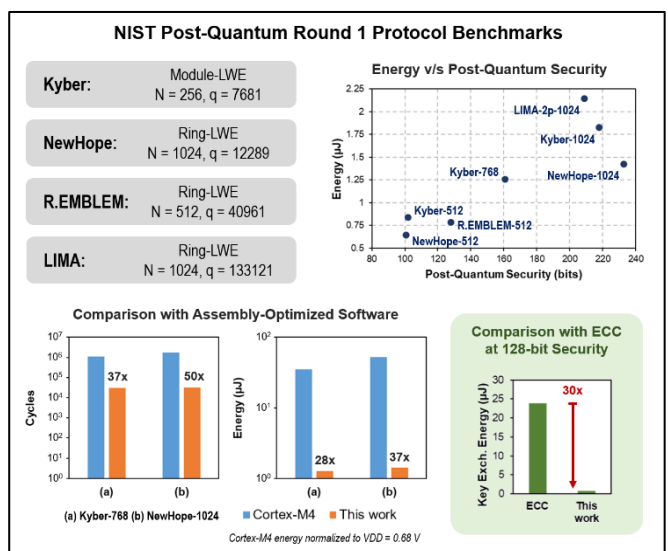

Fig. 5: Configurability of the lattice cryptography processor along with NIST Round 1 post-quantum protocol benchmarks.

Fig. 5 demonstrates the configurability of our processor by benchmarking NIST Round 1 post-quantum protocols such as Kyber [6], NewHope [7], R-EMBLEM [8] and LIMA [9]. Our hardware can be configured for polynomials of length ($N$) 64 to 2048, modulus $q$ up to 24 bits, and discrete distributions with varying standard deviations, thus allowing the processor to tune the security level to provide energy scalability. When executing the Kyber-768 and NewHope-1024 key exchange schemes, our design is respectively 28x and 37x more energy-efficient than Cortex-M4 software, after accounting for voltage scaling. Moreover, post-quantum key exchange using our processor is 30x more energy-efficient than state-of-the-art pre-quantum ECC-based key exchange [10] at the same pre-quantum security level.

| | Cortex-M4 | ISSCC'15 [1] [a] | CICC'18 [4] | This work |
|---|---|---|---|---|
| Technology | – | 130 nm | 40 nm | 40 nm |
| Supply Voltage (V) | 3.0 | 1.2 | 0.9 | 0.68 – 1.1 |
| Frequency (MHz) | 100 | 500 | 300 | 12 – 72 |
| Total Area (mm$^2$) | – | – | 2.05 | **0.28** |
| Logic Gates | – | – | – | 106k |
| Supported Lattice Crypto Primitives | All (Software) | Ring-LWE | Ring-LWE | **Ring-LWE and Module-LWE** |
| Supported Lattice Parameters | All (Software) | N: 256 q: 7681 | N: 64 to 2048 q: 32-bit config. | N: 64 to 2048 q: 24-bit config. |
| NTT Performance (N = 256, q = 7681) | | | | |
| NTT Cycles | 22031 | 1700 [b] | 160 | 1288 [b] |
| NTT Energy (nJ) | 13.5 x 10$^3$ | – | 31 | **63.4** |
| Binomial Sampling Performance (N = 512, q = 12289) | | | | |
| Sample Cycles | 155872 | – | 3704 [c] | 1009 [d] |
| Sample Energy (nJ) | 95.8 x 10$^3$ | – | 1250 | **44.4** |

[a] Post-synthesis data reported for [1]
[b] NTT cycles in [1] and this work include the multiplication of coefficients with powers of $\psi$
[c] [4] uses ChaCha20 as PRNG for sampling
[d] This work uses SHAKE-256 as PRNG for sampling and cycles include the generation of pseudo-random bits

Fig. 6: Comparison with Cortex-M4 software and hardware lattice cryptography accelerators.

Fig. 6 compares this work with software implementation on ARM Cortex-M4 as well as previous work in custom hardware design for lattice-based cryptography. The proposed single-port RAM-based NTT architecture makes our design more area-efficient than [4]. Although the use of multiple parallel butterflies can reduce NTT energy [4], we have used a single butterfly since NTT is only a small fraction of the total computation. An energy-efficient SHA-3 core along with our fast sampling architecture provides 28x energy savings in binomial sampling compared to [4]. This work also demonstrates complete lattice-based protocols, while achieving more than an order of magnitude improvement in energy-efficiency over software.

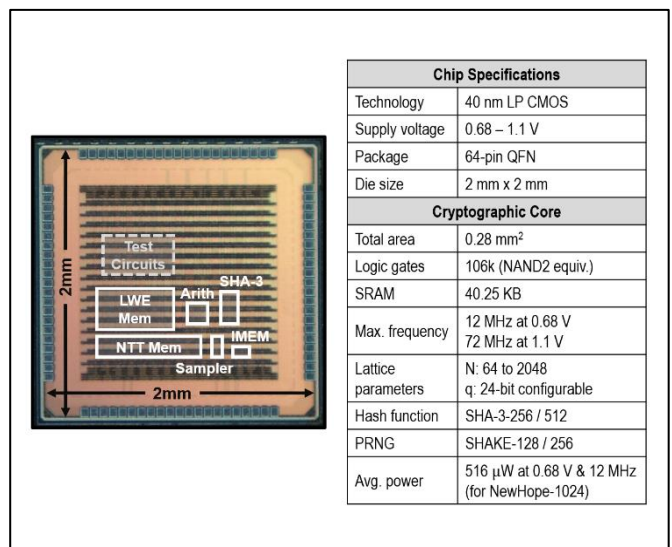

Fig. 7: Chip micrograph and performance summary.

The chip was fabricated in a 40nm LP CMOS process and supports voltage scaling from 1.1V down to 0.68V. All hardware measurements are reported at 12MHz and 0.68V. Our lattice cryptography processor occupies 106k NAND Gate Equivalents (GE) and uses 40.25KB of SRAM. It has an average power of 516 µW when performing the NewHope post-quantum key exchange. Through architectural and algorithmic optimizations, this work demonstrates practical hardware-accelerated quantum-resistant lattice-based cryptographic protocols that can be used to secure resource-constrained IoT devices of the near future.

*Acknowledgements:*

The authors would like to thank Texas Instruments for funding this work, and the TSMC University Shuttle Program for chip fabrication support.

*References:*

[1] I. Verbauwhede et al., "Circuit Challenges from Cryptography," *IEEE ISSCC*, pp. 428-429, 2015.
[2] T. Oder et al., "Practical CCA2-Secure and Masked Ring-LWE Implementation," *IACR Transactions on CHES*, pp. 142-174, 2018.
[3] S. Gueron et al., "Speeding up R-LWE Post-Quantum Key Exchange," *IACR Cryptology ePrint Archive*, Report 2016/467, 2016.
[4] S. Song et al., "LEIA: A 2.05mm2 140mW Lattice Encryption Instruction Accelerator in 40nm CMOS," *IEEE CICC*, pp. 1-4, 2018.
[5] M. C. Pease, "An Adaptation of the Fast Fourier Transform for Parallel Processing," *Journal of the ACM*, pp. 252-264, 1968.
[6] J. Bos et al., "CRYSTALS – Kyber: A CCA-Secure Module-Lattice-Based KEM," *IEEE EuroS&P*, pp. 353-367, 2018.
[7] E. Alkim et al., "Post-Quantum Key Exchange – a New Hope," *USENIX Security*, pp. 327–343, 2016.
[8] M. Seo et al., "EMBLEM – Error-blocked Multi-Bit LWE-based Encapsulation Mechanism," *NIST PQC Round 1*, 2018.
[9] N. Smart et al., "LIMA – A PQC Encryption Scheme," *NIST PQC Round 1*, 2018.
[10] U. Banerjee et al., "An Energy-Efficient Reconfigurable DTLS Cryptographic Engine for End-to-End Security in IoT Applications," *IEEE ISSCC*, pp. 42-44, 2018.